\documentclass[fleqn,final]{slides}
\usepackage{german}
\usepackage[latin1]{inputenc}
\usepackage{vortrag}
\title{UrQMD at RHIC Energies}
\author{\vskip2cm
\color{blue}Marcus Bleicher\\
\vspace*{2cm}
Institut f\"ur Theoretische Physik\\
Goethe-Universit\"at \\
Frankfurt am Main, Germany}
\date{\small\textcolor{green}{May 1999}}

\begin{document}
\maketitle

\begin{slide}
\enlargethispage*{2\baselineskip}
\headline{Collision Spectra}
\bi
\item Center of mass energy of individual collisions
\ei
\figcenter{ws}
\bi
\item Final state is dominated by \textcolor{red}{meson-meson} \\
and \textcolor{red}{meson-baryon} reaction,
\item Fraction of \textcolor{red}{high energy} baryon-baryon \\
reactions is extremely \textcolor{red}{small}
\ei 
\end{slide}

\begin{slide}
\enlargethispage*{2\baselineskip}
\headline{Baryon Rapidity Distribution}
\bi
\item Net protons, net $\Lambda+\Sigma$, net $\Xi$, net $\Omega$.
\ei
\figcenter{rap_netp}
\bi
\item Rapidity distribution of all net baryons shows a \textcolor{red}{dip} 
\item Baryon to anti-baryon ratio at midrapidity: 3/1
\ei 
\end{slide}

\begin{slide}
\enlargethispage*{2\baselineskip}
\headline{Particle Production}
\bi
\item Distributions of all Pions, charged Pions and Kaons 
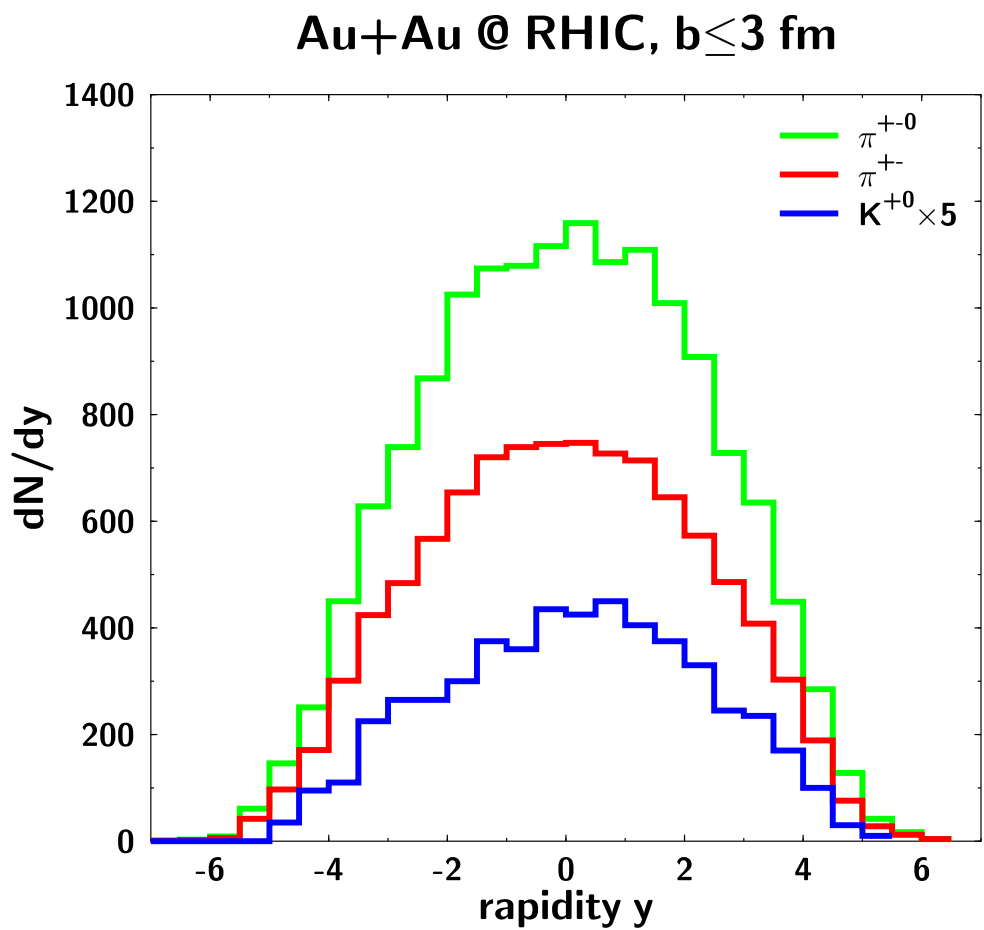
\ei
\figcenter{dndy}
\bi
\item More than \textcolor{red}{1100 Pions} at midrapidity,
\item Approx. \textcolor{red}{750 charged Pions} at midrapidity
\item Approx. \textcolor{red}{80 Kaons} at midrapidity
\ei 
\end{slide}

\begin{slide}
\enlargethispage*{2\baselineskip}
\headline{Creation of transverse expansion}
\bi
\item Mean $p_\perp$ as function of rapidity 
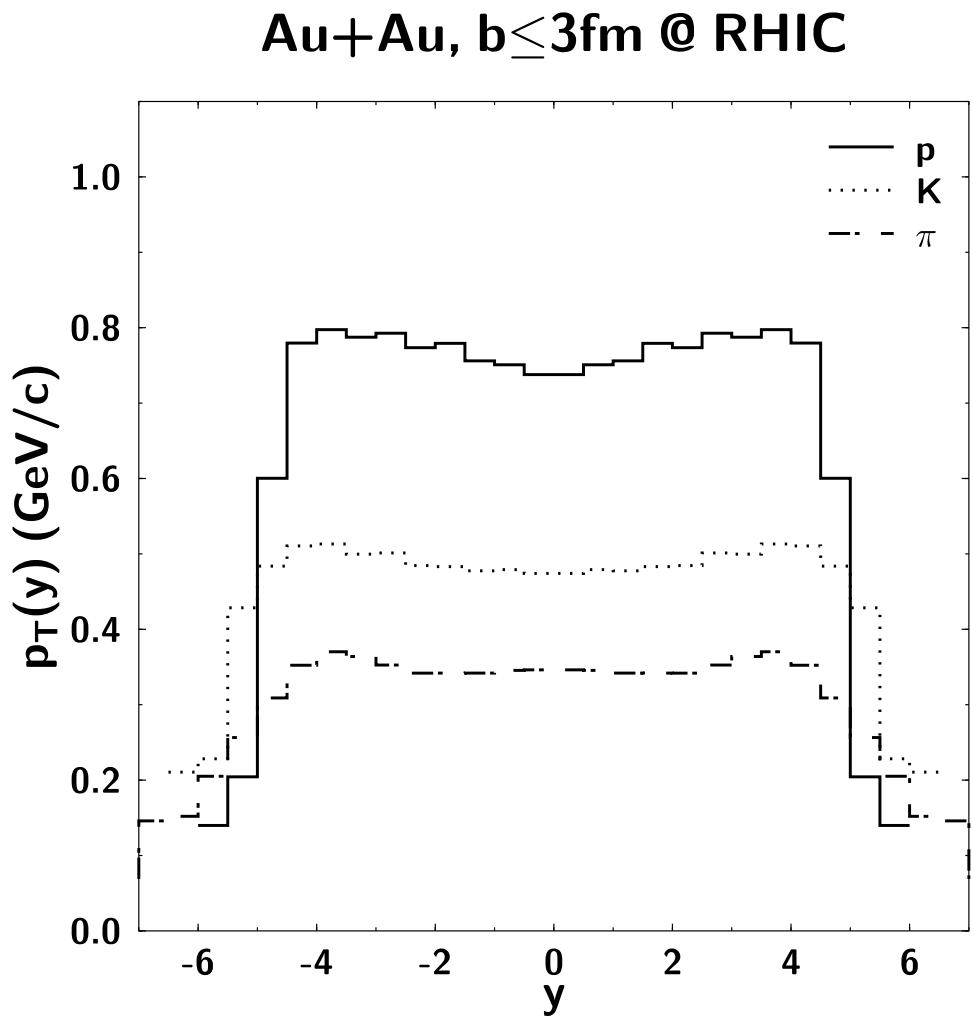
\ei
\figcenter{rhic_mpt}
\bi
\item Remnants of \textcolor{red}{Sea-Gull effect} visible in the Proton distribution (Dip at $y=0$)
\item \textcolor{red}{Plateau} in the transverse momentum distribution of newly produced particles
\ei 
\end{slide}

\begin{slide}
\enlargethispage*{2\baselineskip}
\headline{Transverse momentum spectra}
\bi
\item Distributions of Protons and Pions at midrapidity 
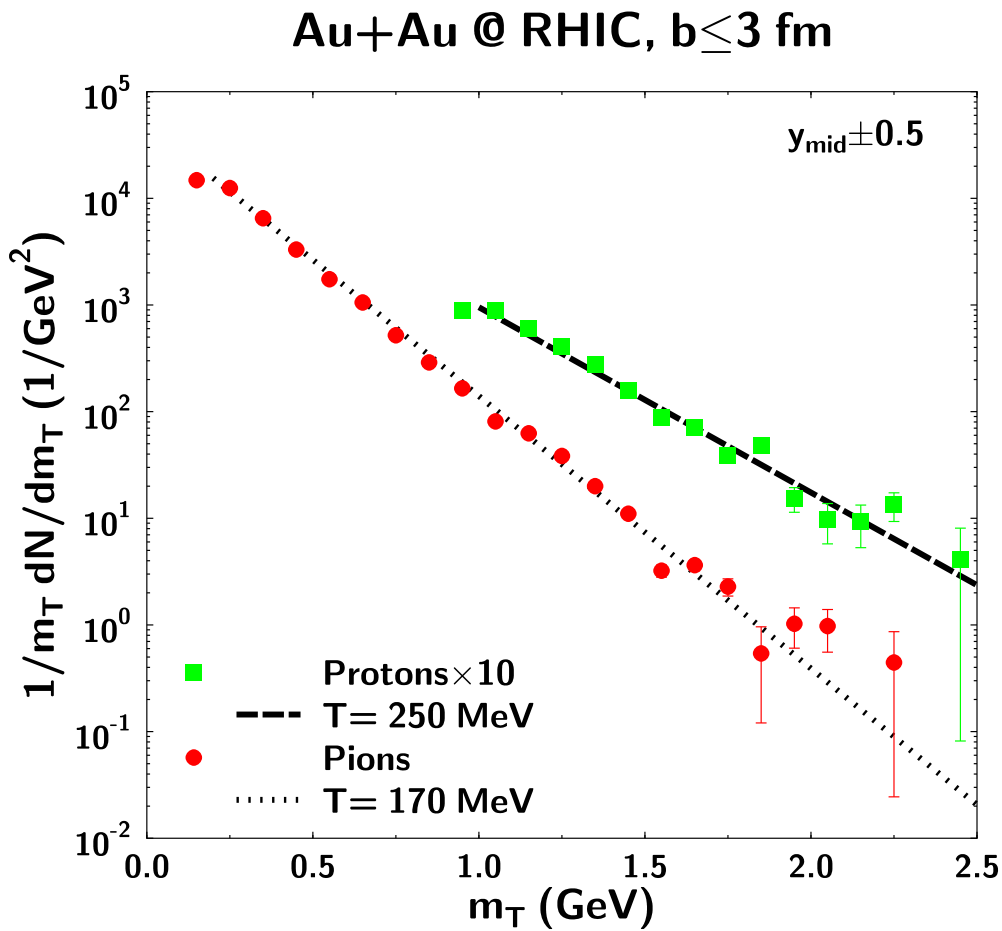
\ei
\figcenter{mt}
\bi
\item Slopes at freeze-out are \textcolor{red}{comparable to SPS} (Pb+Pb)
\item Inv. slopes increase with particle mass
\ei 
\end{slide}

\end{document}